# Quantum mass corrections for affine Toda solitons

N. J. MacKay[1]

*Department of Applied Mathematics and Theoretical Physics, Silver Street, Cambridge, CB3 9EW, U. K.*

and

G. M. T. Watts[2]

*CEA, SPhT, CE-SACLAY, 91191 Gif-sur-Yvette CEDEX, France.*

**ABSTRACT**

We calculate the first quantum corrections to the masses of solitons in some imaginary-coupling affine Toda theories using the semi-classical method of Dashen, Hasslacher and Neveu. The theories divide naturally into those based on the simply-laced, the twisted and the untwisted non-simply-laced algebras. We find that the classical relationships between soliton and particle masses found by Olive *et al.* persist for the first two classes, but do not appear to do so naively for the third.

[1]E-mail address: `n.j.mackay@damtp.cambridge.ac.uk`
[2]E-mail address: `g.m.t.watts@damtp.cambridge.ac.uk`

# 1 Introduction

In [1, 14] Olive *et al.* found a complete set of soliton solutions for the affine Toda field theories. They also pointed out some intriguing relations between particle and soliton masses in different theories. This has echoes of similar conjectures in four-dimensional gauge theories, as first suggested by Olive and Montonen, and now followed up by Seiberg and Witten. The results in Toda theory are classical results, and it is obviously of interest to see whether they persist when the theory is quantized.

To find the quantum corrections to the soliton masses we use the semi-classical method of Dashen *et al.* [2, 3] which was applied to the $a_n^{(1)}$ theories by Hollowood [4]. It is unclear whether this method can be justified within the path integral framework, since it is not clear over which configurations one should sum, but the results Hollowood found seem quite reliable when checked against the results of other non-perturbative methods, such as the S-matrix bootstrap and the method of non-local charges.

The layout of the paper is as follows: In section 2 we recall the known facts about the classical imaginary coupling Toda theories. In section 3 we recall how to calculate the quantum corrections to the particle and soliton masses. In section 4 we give the results for the simply-laced algebras. In section 5 we give the results for the twisted and non-simply-laced untwisted theories. In section 6 we give some comments on the calculations. In appendix A we relate the results of Dorey and Fring and Olive which are relevant to our discussions. In appendix B we give the results for the particle mass corrections in simply-laced theories. In appendix C we list some data associated with the $e_n^{(1)}$ theories.

## 2 Classical imaginary-coupling Toda theory

An affine Toda field theory is a theory of scalar fields in two dimensions with exponential interactions. There is an affine Toda field theory associated with each affine Lie algebra as follows: if we denote the simple roots of the affine Lie algebra by $\alpha_a$, $1 \leq a \leq n$, the lowest root by $\alpha_0$ and the fields by an $n$-dimensional vector $\phi$, then the Lagrangian density is

$$\mathcal{L} = \frac{1}{2}\partial^\mu\phi\cdot\partial_\mu\phi - \frac{m^2}{\beta^2}\sum_a n_a\left[\exp(\beta\alpha_a\cdot\phi) - 1\right] , \qquad (2.1)$$

where $\beta$ and $m$ are coupling constants[1] and $n_a$ are numbers chosen so that $\phi = 0$ is the minimum of the potential. We choose the longest root to have length squared 2, and the $n_a$ to be positive integers such that $\sum n_a\alpha_a = 0$ and $\sum n_a = kh$, where $h$ is the Coxeter number and $k$ the twist of the affine algebra. The different affine algebras and their simple roots are encoded in Dynkin diagrams, which we give in table 1, along with various Lie algebraic data.

For a long time affine Toda field theories were only studied with the coupling constant $\beta$ real (henceforth we shall refer to these as 'real-coupling theories'). As quantum field theories these are theories of (rank $g$) scalar particles. The presence of higher spin conserved quantities in the quantum as well as classical theory (see e.g. [5, 6]) implies that the scattering preserves individual particle momenta, and that the S–matrix factorizes on the two-particle scatterings (see [7–10] for details).

The theories with imaginary coupling constant have a very different spectrum, as can be seen from the sine-Gordon model (henceforth we shall refer to these as 'imaginary-coupling' theories). The potential is periodic and so there are finite energy soliton solutions which interpolate between different vacua. There are also 'breather' solutions which appear as bound states of solitons. For affine Toda theories other than the sine-Gordon model there are difficulties arising because for real values of the fields $\phi$ the potential is not in general real. This means that starting from a real configuration the field will become complex, and so it is hard to see how any 'physical' observables will remain real. The potentials have many stationary points with real values of $\phi$, and one can try to find 'soliton' solutions which interpolate between these stationary points. Hollowood was able to construct such soliton solutions for the $a_n^{(1)}$ theories for which, although the energy density was not real, the energy itself was [11].

This result was succeeded by many others on classical soliton spectra of imaginary coupling affine Toda theories, but the most complete analysis has been given by Olive *et al.*

[1]Note that here we choose the prefactor to the potential to be $m^2/\beta^2$ whereas Olive at al. usually choose this to be $2\mu^2/\beta^2$.

in the series of papers [1, 12–15]. They found that the (classical) soliton masses in simply-laced and twisted theories were in the same ratio as the corresponding (classical) particle masses, whereas the soliton masses for the untwisted affine Lie algebras $g^{(1)}$ were in the same ratios as the particle masses in the theory based on $(g^\vee)^{(1)}$ where the Dynkin diagram of the finite Lie algebra $g$ is that of $g^\vee$ with the arrows reversed. (We shall call this 'Lie duality', in distinction to the 'affine duality' in the real-coupling quantum theory which related theories obtained by reversing the arrows of the Dynkin diagram of the affine Lie algebra.) Thus the masses of the particles in the affine Toda theory based on $b_n^{(1)}$ are in the same ratio as the masses of the solitons in the theory based on $c_n^{(1)}$, and vice versa, whereas for all other theories the particle and soliton mass ratios are equal. For reference we list the classical particle and soliton masses in table 2.

It is obviously an interesting problem to repeat Hollowood's calculation of the quantum soliton mass corrections for the theories other than $a_n^{(1)}$, as has already been performed for $c_2^{(1)}$ in [16]. In this paper we carry out a systematic survey of the other theories.

Table 1: Dynkin diagrams

| | | $h$ | $h^\vee$ |
|---|---|---|---|
| $a_n^{(1)}$ | 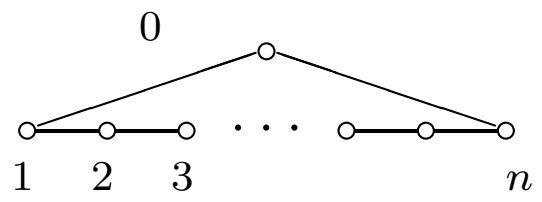 | $n+1$ | $n+1$ |
| $d_n^{(1)}$ | 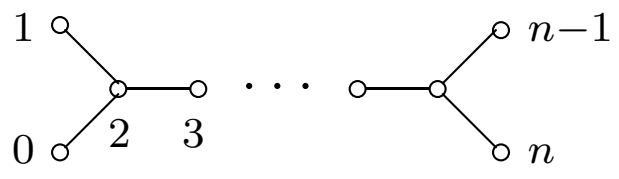 | $2n-2$ | $2n-2$ |
| $e_6^{(1)}$ | 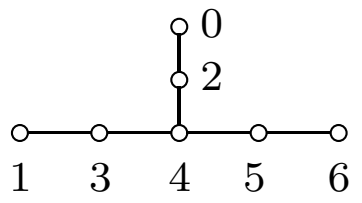 | 12 | 12 |
| $e_7^{(1)}$ | 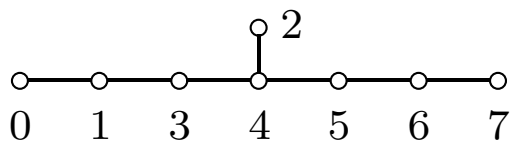 | 18 | 18 |
| $e_8^{(1)}$ | 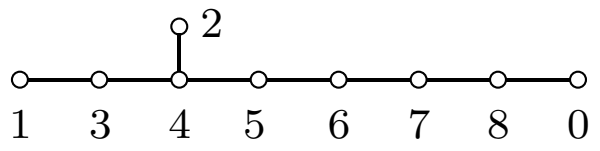 | 30 | 30 |
| $a_{2n}^{(2)}$ | 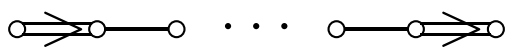 | $2n+1$ | $2n+1$ |
| $a_{2n-1}^{(2)}$ | 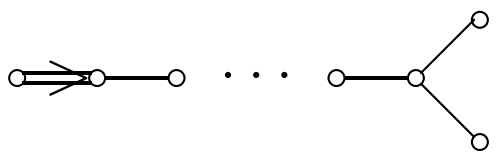 | $2n-1$ | $2n$ |
| $d_{n+1}^{(2)}$ | 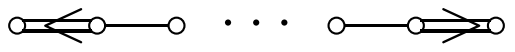 | $n+1$ | $2n$ |
| $d_4^{(3)}$ | 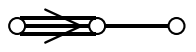 | 4 | 6 |
| $e_6^{(2)}$ | 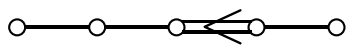 | 9 | 12 |
| $c_n^{(1)}$ | 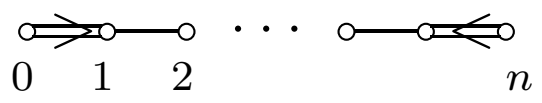 | $2n$ | $n+1$ |
| $b_n^{(1)}$ | 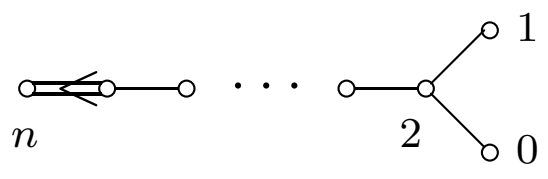 | $2n$ | $2n-1$ |
| $f_4^{(1)}$ | 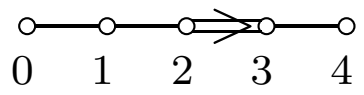 | 12 | 9 |
| $g_2^{(1)}$ | 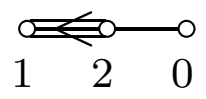 | 6 | 4 |

Table 2: Classical particle and soliton spectra

| | Particle masses | Soliton masses |
|---|---|---|
| $a_n^{(1)}$ | $m_a = 2m\sin\left(\frac{a\pi}{h}\right) \quad a = 1..n$ | $M_a = -\frac{2h}{\beta^2} m_a$ |
| $d_n^{(1)}$ | $m_a = 2\sqrt{2}m\sin\left(\frac{a\pi}{h}\right) \quad a = 1..n-2$<br>$m_n = m_{n-1} = \sqrt{2}m$ | $M_a = -\frac{2h}{\beta^2} m_a$ |
| $a_{2n}^{(2)}$ | $m_a = 2\sqrt{2}m\sin\left(\frac{a\pi}{h}\right) \quad a = 1..n$ | $M_a = -\frac{4h}{\beta^2} m_a$ |
| $a_{2n-1}^{(2)}$ | $m_a = 2\sqrt{2}m\sin\left(\frac{a\pi}{h}\right) \quad a = 1..n-1$<br>$m_n = \sqrt{2}m$ | $M_a = -\frac{4h}{\beta^2} m_a$<br>$M_n = -\frac{4h}{\beta^2} m_n$ |
| $d_{n+1}^{(2)}$ | $m_a = 2\sqrt{2}m\sin\left(\frac{a\pi}{2h}\right) \quad a = 1..n$ | $M_a = -\frac{4h}{\beta^2} m_a$ |
| $d_4^{(3)}$ | $m_L = \sqrt{2}\sqrt{3-\sqrt{3}}\,m$<br>$m_H = \sqrt{2}\sqrt{3+\sqrt{3}}\,m$ | $M_a = -\frac{6h}{\beta^2} m_a$ |
| $c_n^{(1)}$ | $m_a = 2m\sin\left(\frac{a\pi}{h}\right) \quad a = 1..n$ | $M_a = -\frac{4h}{\beta^2} m_a$<br>$M_n = -\frac{2h}{\beta^2} m_n$ |
| $b_n^{(1)}$ | $m_a = 2\sqrt{2}m\sin\left(\frac{a\pi}{h}\right) \quad a = 1..n-1$<br>$m_n = \sqrt{2}m$ | $M_a = -\frac{2h}{\beta^2} m_a$<br>$M_n = -\frac{4h}{\beta^2} m_n$ |
| $g_2^{(1)}$ | $m_1 = \sqrt{2}m,$<br>$m_2 = \sqrt{6}m$ | $M_1 = -\frac{6h}{\beta^2} m_1$<br>$M_2 = -\frac{2h}{\beta^2} m_2$ |

For $f_4^{(1)}$ see subsection 5.2.3; for $e_6^{(2)}$ see subsection 5.1.3; for $e_n^{(1)}$ see [8] and [1, 14].

# 3 Methods of quantization

For real-coupling Toda theories, the spectrum consists of (rank($g$)) massive particles, and one can use standard Feynman diagram techniques to calculate mass corrections, S-matrix elements and so on. One very interesting conjecture is that the strong-coupling behaviour of a quantum affine Toda theory related to a given affine Lie algebra $\hat{g}$ is identical to the weak-coupling behaviour of the affine Toda theory related to $\hat{g}^\vee$ (where the Dynkin diagram of $\hat{g}^\vee$ is obtained by reversing the arrows on the Dynkin diagram of $\hat{g}$) (this has been checked by calculating S-matrix elements and conserved currents at the one-loop level, comparing these with conjectured formulae for the S-matrix elements, and by numerical simulation, see e.g. [6, 9, 17].) For diagrams which are unchanged by reversing the arrows the ratios of the particle masses are not altered by leading-order quantum corrections.

For imaginary-coupling Toda theories, one has to consider the soliton sectors as well. The sine-Gordon theory, which is well-behaved, has been quantized by means of the quantum inverse scattering method, but for the other theories, in addition to the immense complexity of any calculations, it is not possible to find a pseudovacuum to which to apply the algebraic Bethe ansatz. Instead we shall adopt the method used by Hollowood [4]. To calculate the quantum mass corrections for the solitons in the $a_n^{(1)}$ theories, he used the semi-classical method of Dashen et al. [2, 3] and, proceeding formally, found that these corrections were real, and that the ratios of the soliton masses were unchanged by the quantum corrections. Although it is hard to see how this method can be derived from a naive path integral quantization if the action is complex, the results he found for the particles and soliton masses were consistent with S-matrices he conjectured [18], which also exhibited the generic non-unitarity one might expect.

We now review the methods to calculate particle and soliton mass corrections in a little more detail.

### 3.1 Particle mass corrections

In real-coupling Toda theory the only excitations are the fundamental particles and one can use standard Feynman diagram techniques to calculate masses of the particles to any desired order in perturbation theory.

The formula for the 1-loop mass corrections is very simple [8],

$$\delta m_a^2 = \sum_{(b,c)\to a} -\frac{C_{abc}^2}{4\pi\delta}\ \tan^{-1}_{[0,\pi)}\left(\frac{|\delta|}{m_b^2+m_c^2-m_a^2}\right)\ , \tag{3.1}$$

where the sum is over all ordered pairs of particles $(b, c)$ which have a 3-pt coupling to the particle $a$. $C_{abc}$ is the three point coupling found from the Lagrangian,

$$\delta^2 = 2(m_a^2 m_b^2 + m_b^2 m_c^2 + m_c^2 m_a^2) - m_a^4 - m_b^4 - m_c^4\ ,$$

and the inverse tangent takes values in $[0, \pi)$.

These calculations have already been performed for all Toda theories [8, 9, 24] and we simply restate them in our tables for convenience. The only claim for originality we make for the particle mass corrections is the observation in appendix B that the simply-laced results can be obtained using algebraic means.

## 3.2 Soliton mass renormalization in 1+1 dimensional field theory

We recall the semi-classical method of Dashen, Hasslacher and Neveu [2, 3].

To treat the quantum theory in a soliton background one considers the Toda field as the classical soliton $\phi_0$ plus a quantum perturbation. To leading order the quantum perturbation can be expanded in a set of harmonic oscillators, one for each of the bounded solutions of the linearized equation of motion in the soliton background. This linearized equation of motion takes the form of a Schrödinger equation,

$$\left( -\frac{\mathrm{d}^2}{\mathrm{d}x^2} + \left.\frac{\partial^2 V}{\partial \phi^2}\right|_{\phi_0} \right) \delta\phi = \omega^2 \delta\phi \ . \tag{3.2}$$

For the imaginary-coupling Toda theories (as with some other theories; see [2, 3]) it is possible to find all the solutions exactly. Of the bounded solutions to (3.2) there are a continuum of reflectionless 'scattering' solutions which are asymptotically plane waves,

$$\delta\phi(x) \sim \begin{cases} \xi \exp(\mathrm{i}kx) & x \to -\infty \\ \xi \exp(\mathrm{i}kx + \mathrm{i}\delta(k)) & x \to \infty \end{cases} \tag{3.3}$$

(where the 'phase shift' $\delta(k)$ is a constant, possibly complex), and a further set which we call the 'discrete modes'. The quantum state corresponding to a single soliton is taken as the vacuum state for these oscillators, and the mass of the soliton is the difference of the expectation values of the Hamiltonian in this soliton state and in the true vacuum. This is still infinite, and to find a finite expression one must also consider renormalization of $m$; when expressed in terms of the renormalized mass, the soliton masses are finite.

The sum over the ground state energies of the oscillators corresponding to the continuous set of scattering states needs to be regulated. Clearly in eqn. (3.3) the vector $\xi$ must be an eigenvector of the classical mass-squared matrix $M_{ij} = \partial^2 V / \partial\phi_i \partial\phi_j \mid_{\phi=0}$ corresponding to some particle of species $a$ say. Consequently we can consider the solutions to eqn. (3.3) as free particles of type $a$ moving in the background of a soliton of type $b$, in which case let us denote the 'phase shift' in eqn. (3.3) by $\delta_{ab}(k)$. If the $\delta_{ab}(k)$ are real[2], then we can consider the soliton in a large box and impose periodic boundary conditions on $\delta\phi(x)$, in which case, taking the size of the box to infinity, we find the shift in the mass of soliton $b$ to be, neglecting terms of order $\beta^2$,

$$\begin{aligned} \delta M_b &= {}_{\text{soliton}}\langle 0 | \, H \, | 0 \rangle_{\text{soliton}} \ - \ {}_{\text{vac}}\langle 0 | \, H \, | 0 \rangle_{\text{vac}} \ - \ M_b^{\text{classical}} \\ &= \sum_{\substack{\text{Discrete} \\ \text{modes}}} \frac{\omega}{2} - \sum_a \int_{-\infty}^{\infty} \frac{dk}{4\pi} \frac{k \delta_{ab}(k)}{\sqrt{k^2 + m_a^2}} \end{aligned}$$

[2]This is the case in theories which contain only real particles, i.e. all except $a_n^{(1)}$, $d_{2n+1}^{(1)}$ and $e_6^{(1)}$. For the rest we should use instead the regularization suggested by Hollowood. When expressed in terms of the variables $\epsilon_{ab}^p, \alpha_{ab}^p$ of eqn. (3.5), the results will be the same.

$$= \sum_{\substack{\text{Discrete}\\ \text{modes}}} \frac{\omega}{2} - \frac{1}{4\pi}\left[\sum_a \delta_{ab}(k)\,|k|\right]_{-\infty}^{\infty} + \sum_a \int_{-\infty}^{\infty} \frac{dk}{4\pi}\sqrt{k^2+m_a^2}\frac{\mathrm{d}\delta_{ab}}{\mathrm{d}k} \; . \qquad (3.4)$$

In all the affine Toda theories the phase shift for particle type $a$ in background $b$ has the general form

$$\exp \mathrm{i}\delta_{ab}(k) = \prod_p (\mathrm{i}k - \alpha_{ab}^p)^{\epsilon_{ab}^p} \; , \quad \sum_p \epsilon_{ab}^p = 0 \; . \qquad (3.5)$$

Given this form, we can then see that the boundary term is

$$-\frac{1}{4\pi}\left[\sum_a \delta_{ab}(k)\,|k|\right]_{-\infty}^{\infty} = -\sum_{a,p} \frac{1}{2\pi}\,\epsilon_{ab}^p\,\alpha_{ab}^p \; ,$$

and the integral is

$$\sum_a \int_{-\infty}^{\infty} \frac{dk}{4\pi}\sqrt{k^2+m_a^2}\frac{\mathrm{d}\delta_{ab}}{\mathrm{d}k}$$

$$= -\sum_a \left(\Delta_a \sum_p \epsilon_{ab}^p \alpha_{ab}^p\right) - \sum_{a,p} \int_{-\infty}^{\infty} \frac{dk}{4\pi} \frac{\epsilon_{ab}^p \alpha_{ab}^p (m_a^2 - \alpha_{ab}^{p\;2})}{(k^2 + \alpha_{ab}^{p\;2})\sqrt{k^2+m_a^2}} \; , \qquad (3.6)$$

where the divergences are contained in

$$\Delta_a = \int_{-\infty}^{\infty} \frac{dk}{4\pi} \frac{1}{\sqrt{k^2+m_a^2}} \; . \qquad (3.7)$$

As mentioned before, to get a finite answer we need to express the results in terms of the renormalized mass, which is found as follows. The classical masses squared $m_a^2$ are the eigenvalues of

$$M^2 = m^2 \sum_i n_i \alpha_i \alpha_i \; . \qquad (3.8)$$

If we denote the renormalized mass parameter by $m_R$ and put

$$\phi_R = \phi + \delta\phi \; ,$$

where $\delta\phi$ is a constant required to leave $\phi_R = 0$ as the minimum of the (normal-ordered) potential, then these are related to the bare parameters by

$$m_R^2 : e^{\beta\alpha_i\cdot\phi_R} := m^2 e^{\beta\alpha_i\cdot\phi} \; ,$$

for each simple root $\alpha_i$, where $:\;\;:$ represents normal ordering using the free massive propagator, and hence by

$$m_R^2\, e^{\beta\alpha_i\cdot\delta\phi} = m^2 e^{\frac{1}{2}\beta^2\langle(\alpha_i\cdot\phi)^2\rangle} \; . \qquad (3.9)$$

If we multiply the expressions in eqn. (3.9) each to the power of $n_i$ and remember that $\sum n_i = kh$ where $k$ is the twist and $h$ is the Coxeter number of the affine Lie algebra, and also that the mass matrix is given by eqn. (3.8), we obtain

$$m = m_R\left(1 - \frac{\beta^2}{4hk}\sum \frac{m_a^2}{m^2}\Delta_a\right) \; , \qquad (3.10)$$

where $\Delta_a$ is as in (3.7).

The final expression for the mass shift, using (3.10) and (3.4), is

$$\delta M_b = \sum_a \left\{ -\frac{M_b \beta^2}{4hk} \frac{m_a^2}{m^2} - \sum_p \epsilon_{ab}^p \alpha_{ab}^p \right\} \Delta_a \tag{3.11}$$

$$- \sum_{a,p} \frac{1}{2\pi} \epsilon_{ab}^p \alpha_{ab}^p \tag{3.12}$$

$$- \sum_{a,p} \int_{-\infty}^{\infty} \frac{dk}{4\pi} \frac{\epsilon_{ab}^p \alpha_{ab}^p (m_a^2 - {\alpha_{ab}^p}^2)}{(k^2 + {\alpha_{ab}^p}^2)\sqrt{k^2 + m_a^2}} \tag{3.13}$$

$$+ \sum_{\substack{\text{Discrete} \\ \text{modes}}} \frac{1}{2}\omega\ . \tag{3.14}$$

Clearly we need the term (3.11) to vanish for the answer to be finite. This is indeed the case for all the affine Toda theories.

# 4 The results for the simply-laced theories

## 4.1 Particle masses

For the simply-laced theories, Braden et al. [8] found the simple answer

$$\frac{\delta m_a^2}{m_a^2} = -\frac{\beta^2}{4h}\cot\theta \ . \tag{4.1}$$

where for the simply-laced theories $\theta$ is defined to be

$$\theta = \frac{\pi}{h} \ . \tag{4.2}$$

In their paper, they proved this in each case in turn.

From Dorey [19] we know that we can associate each particle in a simply-laced Toda theory with an orbit of the Coxeter element in the space of roots. In appendix A we give our conventions for the choice of these orbit representatives based on Dorey [D] and Fring and Olive [FO]. Using techniques developed by Dorey [19–21] and Fring and Olive [22] we prove the result (4.1) for even $h$ in the appendix B. As can be seen, the universal form of the mass shifts implies that the masses remain in the same ratio to 1 loop.

## 4.2 Soliton masses

### 4.2.1 Soliton solutions

From Olive et al. [1] we know that with each orbit there is associated a soliton with mass $M_i$

$$M_i = -\frac{2hm_i}{\beta^2} \ .$$

They give the $m$–soliton solutions of a simply-laced affine Toda theory as

$$\phi = -\frac{1}{\beta}\sum_{j=0}^{\text{rank}} \alpha_j \ \log\tau_j \tag{4.3}$$

$$\tau_j = \langle j| \ \exp(W_{i_1}(z_{i_1})\hat{F}^{i_1}(z_{i_1}))\ldots \ \exp(W_{i_m}(z_{i_m})\hat{F}^{i_m}(z_{i_m})) \ |j\rangle \tag{4.4}$$

where $\hat{F}^i(z_i)$ are generators of the simply-laced affine algebra $\hat{g}$, $i$ labels the orbit of the Coxeter group and hence the soliton species, $|j\rangle$ is a highest weight state corresponding to the $j$th fundamental weight of level[3] $m_j$,

[3] $m_j$ is the lowest integer level at which the $j$th fundamental representation is an allowed representation of $\hat{g}$

$$W(z_i) = q_i \exp(m_i(x\cosh\eta_i + t\sinh\eta_i)) \ , \quad z_i = \mathrm{i}\exp(\ i\theta\frac{(1-c_i)}{2} - \eta_i\ ) \ ,$$

$\eta_i$ is the rapidity of the $i$th soliton, $c_i = \pm 1$ and the $q_i$ are arbitrary complex parameters which determine the topological charge and centre of mass trajectory of the $i$th soliton. We refer the reader to [1, 12–14] for full details concerning the various elements used in this construction and to appendix A for our conventions.

### 4.2.2 The solutions to the linearized equations of motion

It is straightforward to find the exact set of solutions to (3.2) using the same trick as Hollowood in [4]. Since the parameters $q_i$ are arbitrary, the $O(q_2)$ term in a two-soliton solution will satisfy the linearized equations of motion in the first soliton background. By running over all possible perturbing soliton types we find a complete set of solutions $\delta\phi^{ab}$ to (3.2) of the form

$$\delta\phi^{ab} = \sum_{j=0} \alpha_j\ W_a(k,x)\ \delta\tau_j^{ab}(k,x) \tag{4.5}$$

$$\delta\tau_j^{ab}(k,x) = \frac{\langle j|\ \hat{F}^a(z_a)\exp(W_b(z_b)\hat{F}^b(z_b))\ |j\rangle}{\langle j|\ \exp(W_b(z_b)\hat{F}^b(z_b))\ |j\rangle} \ , \tag{4.6}$$

where we take the soliton $b$ to be at rest, and so $W_b(z_b) = \exp(m_b x)$.

Using the results and notation of [12, 13], we can examine the limits $x \to \pm\infty$ of these functions. The limit $x \to -\infty$ is straightforward as $W_b \to 0$, and so

$$\delta\tau_j^{ab}(k,x) \sim_{x\to-\infty} \frac{\langle j|\ \hat{F}^a(z_a)\ |j\rangle}{\langle j|j\rangle} = F_{ja} \ , \tag{4.7}$$

and so

$$\delta\phi^{ab}(k,x) \sim_{x\to-\infty} \xi_a \exp(\mathrm{i}kx - \mathrm{i}\omega t) \ , \tag{4.8}$$

where $\omega^2 - k^2 = m_a^2$ and

$$\xi_a = \sum_{j=0} \alpha_j F_{ja} \tag{4.9}$$

is an eigenvector of the mass-squared matrix of eigenvalue $m_a^2$. Thus as $x \to -\infty$ this appears to be a particle of type $a$ moving in the background of a soliton of type $b$.

As $x \to \infty$ we have (using extensively [14, 15])

$$\delta\tau_j^{ab}(k,x) \sim_{x\to\infty} \frac{\langle j|\ \hat{F}^a(z_a)(\hat{F}^b(z_b))^{m_j}\ |j\rangle}{\langle j|\ (\hat{F}^b(z_b))^{m_j}\ |j\rangle} = X_{ab}(z_a,z_b)F_{ja} \ , \tag{4.10}$$

$$X_{ab}(z_a, z_b) = \prod_{p=1}^{h} (\frac{z_a}{z_b} - \omega^p)^{\gamma_a \cdot \sigma^p(\gamma_b)} \ , \tag{4.11}$$

where $\sigma$ is the coxeter element, $\gamma_i$ are representatives of the coxeter orbits (see App. A), and $\omega = \exp(\, 2\mathrm{i}\theta\, )$. So we see that $\delta\phi^{ab}$ is of the form (3.3) with the phase shift given by

$$\exp(\mathrm{i}\delta_{ab}) = X_{ab}(z_a, z_b) \ , \tag{4.12}$$

if we choose

$$m_a \cosh\eta = \mathrm{i}k \ .$$

Since there are no singularities in $X_{ab}$ for $k$ real, this is sufficient to define a complete set of 'scattering states' and calculate all the contributions to the mass shifts from these states.

### 4.2.3 Using the phase shifts for the simply-laced algebras

Using the expressions for $z_a$, Olive et al. [1] give the simply-laced $X$ factors as

$$\prod_{p=1}^{h} \left( e^{\eta} - \omega^p e^{i\theta \frac{c_i - c_k}{2}} \right)^{\gamma_i \cdot \sigma^p \gamma_k} \ .$$

We can turn this into an expression entirely in $\cosh\eta$ by using the fact that it is even in $\theta$ and multiplying by the expression with $\eta \to -\eta$ and then taking the square root, at the expense of apparently non-integral exponents

$$\begin{array}{l} \prod_{p=1}^{h} \left( \cosh\eta - \cos\theta(2p + \frac{c_i - c_k}{2}) \right)^{\frac{1}{2}\gamma_i \cdot \sigma^p \gamma_k} \ , \\ \prod_{p=1}^{h} \left( \cosh\eta - \cos\theta(2p + u_{ab}) \right)^{\frac{1}{2}\phi_a \cdot w^{-p} \phi_b} \ , \end{array} \tag{4.13}$$

where we give the [FO] and [D] versions. Remembering that $ik = m_a \cosh\eta$, we find $\exp \mathrm{i}\delta_{ab}$ is of the form (3.5) with

$$\epsilon_{ab}^p = \frac{1}{2}\phi_a \cdot w^{-p}\phi_b \ \ , \ \alpha_{ab}^p = m_a \cos(2p + u_{ab})\theta \ . \tag{4.14}$$

With this form of the phase shift, we can now evaluate the boundary term and the integral for the simply-laced theories.

- The (potential) divergence

  Using eqns. (A.2) and (A.3) of appendix A we have the result

  $$\sum_p \phi_a \cdot w^{-p}\phi_b \cos(2p + u_{ab})\theta = \frac{m_a m_b}{m^2} \ ,$$

  and hence the divergence in the soliton mass corrections is cancelled by the mass renormalization and (3.11) is identically zero.

- Boundary terms:

  Putting (4.14) into (3.12) we find

$$
\begin{aligned}
-\sum_a \sum_{p=1}^{h} \frac{1}{2\pi} \epsilon_{ab}^p \alpha_{ab}^p &= -\sum_a \sum_{p=1}^{h} \frac{m_a}{4\pi} \phi_a \cdot w^{-p} \phi_b \cos(2p+u_{ab})\theta \\
&= -\frac{h m_b}{2\pi} \ . \qquad (4.15)
\end{aligned}
$$

- The remaining integral (3.13) now becomes

$$
\begin{aligned}
I &= -\sum_{a,p} \int_{-\infty}^{\infty} \frac{dk}{4\pi} \frac{\epsilon_{ab}^p \alpha_{ab}^p (m_a^2 - \alpha_{ab}^{p2})}{(k^2+\alpha_{ab}^2)\sqrt{k^2+m_a^2}} \\
&= -\sum_{a,p} \frac{m_a}{4\pi} \phi_b \cdot w^{-p} \phi_a \cos(2p+u_{ba})\theta \sin^2(2p+u_{ba})\theta \\
&\times \left| \frac{\tan^{-1}_{(-\pi/2,\pi/2)} \tan(2p+u_{ab})\theta}{\cos(2p+u_{ba})\theta \sin(2p+u_{ba})\theta} \right| \ . \qquad (4.16)
\end{aligned}
$$

Let us assume that $b$ is a 'white' index, and hence $u_{b\circ}=0$, $u_{b\bullet}=1$. Then we also have

$$
\sum_\circ m_\circ \phi_\circ = \left(m\sqrt{2h}\right) a_\circ^1 \ , \ \sum_\bullet m_\bullet \phi_\bullet = \left(m\sqrt{2h}\right) w_{\{\circ\}} a_\bullet^1 \ ,
$$

and if we remember that for $b$ 'white' (A.4)

$$
\left(m\sqrt{2h}\right) \phi_b \cdot w^{-p} w_{\{\circ\}} a_\bullet^1 = 2 m_b \cos(2p+1)\theta \ , \ \left(m\sqrt{2h}\right) \phi_b \cdot w^{-p} a_\circ^1 = 2 m_b \cos 2p\theta \ ,
$$

then we can perform the sums over the black and white roots independently, and recombine them to give

$$
I = -\sum_{p=0}^{2h-1} \frac{m_b}{2\pi} \cos^2 p\theta \sin^2 p\theta \left| \frac{\tan^{-1}_{(-\pi/2,\pi/2)} \tan p\theta}{\cos p\theta \sin p\theta} \right| \ .
$$

If we take $h$ even, then this gives the answer

$$
I = -\frac{m_b}{4} \cot\theta \ .
$$

The case of $h$ odd only occurs for the $a_n^{(1)}$ series, and the (different) result in that case has been found by Hollowood in [4].

### 4.2.4 The discrete modes

The remaining discrete set of solutions to eqn. (3.2) have the property that $\delta\phi_{ab} \to$ const. as $x \to \pm\infty$. From eqn. (4.8) we see that i$k$ must be real and non-negative. The $k = 0$ solutions are already included in the 'scattering states', so we must take i$k > 0$. In this case a necessary condition for $\delta\phi_{ab}$ to be bounded as $x \to \infty$ is that $X_{ab}(z_a, z_b) = 0$.

Let us suppose that this is so, in which case we have

$$\delta\tau_j^{ab}(k,x) \sim_{x\to\infty} m_j W_b(z_b) \frac{\langle j|\,\hat{F}^a(z_a)(\hat{F}^b(z_b))^{m_j-1}\,|j\rangle}{\langle j|\,(\hat{F}^b(z_b))^{m_j}\,|j\rangle}\ . \tag{4.17}$$

It is possible to calculate this explicitly again using the results of [1, 13]. Let us recall the level one vertex operator construction of $\hat{F}^k(z)$ of [1],

$$\begin{aligned} \hat{F}^k(z) &= \tilde{F}_0^k \exp\left(\sum_{M>0} \frac{1}{M}\gamma_k \cdot q([M]) z^M \hat{E}_{-M}\right) \exp\left(\sum_{M>0} \frac{1}{-M}\gamma_k \cdot q([-M]) z^{-M} \hat{E}_M\right) \\ &= \tilde{F}_0^k \hat{F}_<^k \hat{F}_>^k\ , \end{aligned} \tag{4.18}$$

where for each level 1 highest weight vector

$$\tilde{F}_0^k\,|j\rangle = \exp(-2\pi i \lambda_k \cdot \lambda_j)\,|j\rangle\ . \tag{4.19}$$

(The $\hat{E}_M$ generate the principal Heisenberg subalgebra, $[M]$ means $M$ mod $h$, and the vectors $q(a)$ are defined in appendix A ) In eqn. (4.11), $\gamma_a \cdot \sigma^{-p}\gamma_b$ can only take values $0, \pm 1, \pm 2$. For $X_{ab}(z_a, z_b) = 0$ this gives two options,

- $\gamma_a \cdot \sigma^{-p}\gamma_b = 2$. In this case we must have $a = b$ and $p = 0$, $z_a = z_b$ and so $k = 0$. We have already counted this solution.
- $\gamma_a \cdot \sigma^{-p}\gamma_b = 1$.

In this case (which is governed by Dorey's fusing rule) there is a third root $\gamma_c$ such that

$$\gamma_a = \sigma^{-p}(\gamma_b) + \sigma^q(\gamma_c) \tag{4.20}$$

and $z_a = \omega^{-p} z_b$, $z_c = \omega^{-q} z_a$. We also have [23] that $\lambda_a - \lambda_b - \lambda_c \in \Lambda_R$, the root lattice of $g$, so that

$$\tilde{F}_0^a = \tilde{F}_0^b \tilde{F}_0^c$$

and as a result

$$\hat{F}^a(z_a) = \tilde{F}_0^b\,\tilde{F}_0^c\,\hat{F}_<^b\,\hat{F}_<^c\,\hat{F}_>^c\,\hat{F}_>^b = \tilde{F}_0^a\,\hat{F}_<^b\,\hat{F}^c(z_c)\,\hat{F}_>^b\ . \tag{4.21}$$

To use this result, we now recall that the level $x$ construction of $\hat{F}^k(z)$ from [1] is simply the $x$–fold tensor product of some level 1 representations,

$$\hat{F}^k(z)\Big|_{\text{level}\,x} = \hat{F}^k \otimes 1 \otimes \ldots \otimes 1 + \ldots + 1 \otimes 1 \otimes \ldots \otimes \hat{F}^k \ , \tag{4.22}$$

and that the $x$th power of this generator again has a simple vertex operator construction exactly as in eqn. (4.18),

$$\begin{aligned} \frac{\left(\hat{F}^k(z_k)\right)^x}{x!} &= \tilde{F}_0^k \ \exp(\sum_{M>0} \frac{1}{M}\gamma_k \cdot q([M]) z^M \hat{E}_{-M}) \ \exp(\sum_{M>0} \frac{1}{-M}\gamma_k \cdot q([-M]) z^{-M} \hat{E}_M) \\ &= \tilde{F}_0^k \ Y^k(z_k) \ Z^k(z_k) \ , \end{aligned} \tag{4.23}$$

where $Y$ and $Z$ are the notation of [13] and $\tilde{F}_0^k$ again satisfies (4.19) but now for level $x$ highest weight states $|j\rangle$.

So, using eqns. (4.21), (4.22) and (4.23) we find that, when $X_{ab}(z_a, z_b)$ has a simple zero,

$$\hat{F}^a(z_a)\frac{\left(\hat{F}^b(z_b)\right)^{x-1}}{(x-1)!} = \tilde{F}_0^b \ Y^b(z_a) \ \hat{F}^c(z_c) \ Z^b(z_a) \ . \tag{4.24}$$

Taking expectation values in $|j\rangle$,

$$\frac{1}{(m_j-1)!}\langle j| \ \hat{F}^a(z_a)(\hat{F}^b(z_b))^{m_j-1} \ |j\rangle \ = \ \exp(-2\pi i \lambda_b \cdot \lambda_j) \ F_{jc} \tag{4.25}$$

and the next-to-leading term (4.17) is

$$m_j W_b(z_b) \frac{\langle j| \ \hat{F}^a(z_a)(\hat{F}^b(z_b))^{m_j-1} \ |j\rangle}{\langle j| \ (\hat{F}^b(z_b))^{m_j} \ |j\rangle} = F_{jc} W_b(z_b) \ , \tag{4.26}$$

and

$$\delta\phi_{ab} \sim_{x\to\infty} \ \frac{W_a(z_a)}{W_b(z_b)} \ \xi_c \ . \tag{4.27}$$

We find that

$$W_a(z_a) \sim \exp\left(x m_a \cos\left((2p + u_{ab})\theta\right)\right) \ ,$$

and so a necessary and sufficient condition for there to be a bound state solution to (3.2) coming from a particle of type $a$ bound to the soliton $b$ is that

$$\gamma_a = \sigma^{-p}(\gamma_b) + \sigma^q(\gamma_c) \ , \ \ 0 < m_a \cos(2p + u_{ab})\theta \le m_b \ . \tag{4.28}$$

At present we have not found a uniform way to sum these contributions for the simply-laced algebras, and we find case by case for the algebras $a_{2n+1}^{(1)}$, $d_n^{(1)}$ and $e_n^{(1)}$ the result for (3.14)

$$\frac{1}{2} \sum_{\substack{\text{Discrete} \\ \text{modes}}} \omega \ = \ \frac{m_b}{2} \cot\theta \ . \tag{4.29}$$

We do not give details here because the sums are straightforward for the classical algebras and for the exceptional algebras we used Mathematica. The case of $h$ odd is again different, only occurs for the $a_n^{(1)}$ series, and has been found by Hollowood in [4].

## 4.3 Final results for simply-laced algebras

Combining the results we have found for $d_n^{(1)}$ and $e_n^{(1)}$ with the results of Hollowood for $a_n^{(1)}$ we produce the final result (in each case in this table $\theta = \frac{\pi}{h}$).

Table 3: Simply-laced algebra results

$$m_a^{qu} = m_a^{cl}\left(1 \;-\; \frac{\beta^2}{8h}\cot\frac{\pi}{h}\right)$$

$$M_a^{qu} = M_a^{cl}\left(1 \;-\; \frac{\beta^2}{8h}\cot\frac{\pi}{h} + \frac{\beta^2}{4\pi}\right)$$

## 5 The results for the non-simply-laced theories

The remaining affine Lie algebras are all non-simply-laced (that is, the simple roots are not all of the same length) and fall into two classes, the non-simply-laced untwisted algebras and the twisted algebras. For the particles in theories based on these algebras, Braden *et al.* found that the mass shifts are no longer universal; the ratios of the particle masses change to first order in perturbation theory. We list these mass shifts in tables 4 and 5.

Solitons for the theories based on non-simply-laced algebras can be obtained from the simply-laced algebras by using automorphisms ('foldings') of their Dynkin diagrams [25]. Suppose the simply-laced ('parent') diagram has some symmetry: then the equations of motion will also have this symmetry, and any initial data with the symmetry will remain symmetric as they evolve in time. Such a solution will thus also be a solution of the folded theory, based on combinations of roots invariant under the automorphism, which are the simple roots of a non-simply-laced algebra. The procedure is best viewed graphically as a folding-together of the legs of the parent diagram exchanged by the automorphism. Full details can be found in [8], whose notational conventions we follow.

Such foldings can be divided into the indirect and the direct: the former exchange roots which are linked on the parent diagram whilst the latter do not. It turns out that to find all theories it is necessary only to consider the direct foldings. The direct foldings divide into those which involve the extended root and those which do not, *i.e.* those which are also symmetries of the (unextended) diagram of the finite Lie algebra. The former lead to the twisted theories and the latter to the non-simply-laced untwisted theories.

Viewed in this way it is simple to compute the soliton solutions for the non-simply-laced theories, since they are soliton solutions in the parent theory which are invariant under the folding. The only difficulty is in identifying which parent (multi)soliton solutions correspond to the non-simply-laced theory's single solitons. It is also straightforward to find the $X$ factors since they come from a particle traversing the parent (multi)soliton.

The $a_n^{(1)}$ factors are given in [4], the $e_6^{(1)}$ and $e_7^{(1)}$ factors in tables 7 and 8 respectively. The $d_n^{(1)}$ theories split into two classes, $n$ even and $n$ odd, (see [8]). For $n$ even, all the particles are self-conjugate, whereas for $n$ odd the two particles corresponding to the two spinor representations form a conjugate pair. The X-factors, however, have uniform expressions

$$\begin{aligned} X_{a\,b} &= \{\cos(a-b)\theta\}\{-\cos(a+b)\theta\} \\ X_{n\,n} = X_{n-1\,n-1} &= \prod_{p=0}^{[(n-2)/2]} \frac{\cosh\eta - \cos 4p\theta}{\cosh\eta - \cos(4p+2)\theta} \end{aligned}$$

$$X_{n-1\,n} = \prod_{p=1}^{[(n-1)/2]} \frac{\cosh\eta - \cos(4p-2)\theta}{\cosh\eta - \cos 4p\theta}$$
$$X_{n\,a} = X_{n-1\,a} = \{\sin(a\pi/h)\} \ ,$$

where $\theta = \pi/h$ and

$$\{a\} \equiv \frac{\cosh\eta - a}{\cosh\eta + a} \ .$$

For $n$ even, these X-factors all give real phase shifts, whereas for $n$ odd the phase shifts for two spinorial labels are complex.

## 5.1 Twisted theories

For the twisted theories, $a_n^{(2)}$, $d_{n+1}^{(2)}$, $e_6^{(2)}$ and $d_4^{(3)}$, the folding procedure produces a potential with the $n_a$ no longer minimal: if $k$ is the order of the automorphism, then $n_0 = k$ and $\sum n_1 = kh$. The longest root is of squared length 2.

### 5.1.1 Particle masses

The particle masses and three point couplings are given in Braden et al. [8] and it is easy to calculate the mass shifts using this data. We give the results in table 4 along with the soliton mass corrections[4].

### 5.1.2 Soliton solutions

In the twisted cases it turns out that all the single solitons can be obtained by folding single solitons in the parent theory [14, 26]. Specifically,

$a_{2n}^{(2)}$ : The parent theory is $d_{2n+2}^{(1)}$ and only even tensor solitons survive folding.

$a_{2n-1}^{(2)}$ : The parent theory is $d_{2n}^{(1)}$ and only the even tensor solitons and one spinor soliton survive.

$d_{n+1}^{(2)}$ : The parent theory is $d_{n+2}^{(1)}$ and only tensor solitons survive.

$e_6^{(2)}$ : The parent theory is $e_7^{(1)}$ and the solitons which survive correspond to the particles $\{2, 4, 5, 7\}$ in the notation of [8]. The relevant $X$-factors may be extracted from table 8.

[4] We should like to thank H. Braden for sending us some of his results on the twisted algebras which were not published in [8]

$d_4^{(3)}$ : The parent theory is $e_6^{(1)}$ and the solitons corresponding to the particles $L$ and $H$ in the notation of [8] survive. The relevant $X$-factors may be extracted from table 7, and are

$$X_{LL} = \{1\}\{\frac{1}{2}\}\{-\frac{\sqrt{3}}{2}\}\ , \quad X_{HL} = \{\cos\frac{\pi}{12}\}\{\cos\frac{5\pi}{12}\}\ , \quad X_{HH} = \{1\}\{\frac{1}{2}\}\{\frac{\sqrt{3}}{2}\}\ .$$

In each case since the solitons in the parent theory are related to the parent particle masses by a universal factor, this remains the case in the twisted theories. The relation is now

$$M_a = -\frac{2hk}{\beta^2} m_a$$

For the twisted theories the $X$s are simply those of the parent single solitons. This means that we can use these $X$ factors directly to calculate the phase shifts and discrete spectrum of the twisted theories and find the corrections to the soliton masses. We give these in table 4.

Table 4: Twisted theory particle and soliton mass corrections

| Theory | Particles | Solitons |
|---|---|---|
| $a_{2n}^{(2)}$<br>$\theta = \frac{\pi}{h}$ | $\frac{m_a^{\rm qu}}{m_a^{\rm cl}} = \left(1 - \frac{\beta^2}{8kh}\cot\theta\right)$ | $\frac{M_a^{\rm qu}}{M_a^{\rm cl}} = \left(1 - \frac{\beta^2}{8kh}\cot\theta + \frac{\beta^2}{4k\pi}\right)$ |
| $d_{n+1}^{(2)}$<br>$\theta = \frac{\pi}{2h}$ | $\frac{m_a^{\rm qu}}{m_a^{\rm cl}} = \left(1 - \frac{\beta^2}{8kh}\cot\theta + \frac{\beta^2}{4kh^2} a \cot a\theta\right)$ | $\frac{M_a^{\rm qu}}{M_a^{\rm cl}} = \left(\begin{array}{c} 1 - \frac{\beta^2}{8kh}\cot\theta + \frac{\beta^2}{4k\pi}\frac{h^\vee}{h} \\ + \frac{\beta^2}{4kh^2} a \cot a\theta \end{array}\right)$ |
| $a_{2n-1}^{(2)}$<br>$\theta = \frac{\pi}{h}$ | $\frac{m_a^{\rm qu}}{m_a^{\rm cl}} = \left(1 - \frac{\beta^2}{8kh}\cot\theta - \frac{\beta^2}{4kh^2} a \cot a\theta\right)$<br>$\frac{m_n^{\rm qu}}{m_n^{\rm cl}} = \left(1 - \frac{\beta^2}{8kh}\cot\theta\right)$ | $\frac{M_a^{\rm qu}}{M_a^{\rm cl}} = \left(\begin{array}{c} 1 - \frac{\beta^2}{8kh}\cot\theta + \frac{\beta^2}{4k\pi}\frac{h^\vee}{h} \\ - \frac{\beta^2}{4kh^2} a \cot a\theta \end{array}\right)$<br>$\frac{M_n^{\rm qu}}{M_n^{\rm cl}} = \left(1 - \frac{\beta^2}{8kh}\cot\theta + \frac{\beta^2}{4k\pi}\frac{h^\vee}{h}\right)$ |
| $d_4^{(3)}$ | $\frac{m_L^{\rm qu}}{m_L^{\rm cl}} = \left(1 - \frac{\beta^2}{192}(1+\sqrt{3})\right)$<br>$\frac{m_H^{\rm qu}}{m_H^{\rm cl}} = \left(1 - \frac{\beta^2}{192}(5-\sqrt{3})\right)$ | $\frac{M_L^{\rm qu}}{M_L^{\rm cl}} = \left(1 - \frac{\beta^2}{192}(1+\sqrt{3}) + \frac{\beta^2}{8\pi}\right)$<br>$\frac{M_H^{\rm qu}}{M_H^{\rm cl}} = \left(1 - \frac{\beta^2}{192}(5-\sqrt{3}) + \frac{\beta^2}{8\pi}\right)$ |

#### 5.1.3 The $e_6^{(2)}$ results

The $e_6^{(2)}$ theory is obtained by folding the $e_7^{(1)}$ theory. We label the particles in accordance with the labelling of [8]. They have masses (with $\theta = \pi/18$),

$$\begin{array}{lll} m_1^2 &=& 8\sqrt{3}m^2 \sin\theta \ \sin 4\theta \ , \qquad m_2^2 = 8\sqrt{3}m^2 \sin 5\theta \ \sin 2\theta \ , \\ m_3^2 &=& 6m^2 \ , \qquad\qquad\qquad\quad m_4^2 = 8\sqrt{3}m^2 \sin 7\theta \ \sin 8\theta \ . \end{array}$$

We list the particle mass corrections below for reference.

| | $m^{\rm qu}/m^{\rm cl}$ |
|---|---|
| 1 | $1+\frac{\beta^2}{216}(\cot\theta - 2\cot 4\theta) - \epsilon$ |
| 2 | $1+\frac{\beta^2}{216}(3\cot 3\theta - \cot 2\theta) - \epsilon$ |
| 3 | $1+\frac{\beta^2}{216}(2\cot 2\theta - 2\cot 4\theta) - \epsilon$ |
| 4 | $1+\frac{\beta^2}{216}(3\cot 3\theta - 2\cot 4\theta) - \epsilon$ |

,

$$\epsilon = \frac{\beta^2}{432 \sin 4\theta}(4 + \sin\theta + 7\cos 2\theta) \ .$$

We find the soliton mass corrections (using Mathematica) to satisfy the universal rule of eqn. (5.1).

#### 5.1.4 Summary of twisted results

In each case the particle and soliton mass corrections are related by

$$\frac{M_a^{\rm qu}}{m_a^{\rm qu}} = -\frac{2hk}{\beta^2}\left(1 + \frac{\beta^2}{4k\pi}\frac{h^\vee}{h} \ + \ O(\beta^4)\right) = -\frac{h^\vee}{\pi B}\left(1 \ + \ O(\beta^4)\right) \ , \qquad (5.1)$$

where $B$ is the universal $\beta$-dependent function used by Dorey [36] to construct particle $S$-matrices for the real-coupling case. (Note that the $k$ is missing in his case because he takes the longest root to have squared length $2k$.)

### 5.2 Untwisted non-simply-laced theories

#### 5.2.1 Particles

Again the particle masses and three point couplings have been given in Braden *et al.* [8] and we list the mass corrections in table 5.

#### 5.2.2 Soliton solutions

In contrast to the twisted cases, in the untwisted cases some multisoliton solutions of the parent theory must be folded to produce single solitons in the daughter theory. The (classical) soliton masses then form the left Perron-Frobenius eigenvector of the (finite Lie algebra's) Cartan matrix, in contrast to the particle masses which form the right eigenvector. This suggests a duality (which we call 'Lie duality')[1, 12] between solitons and particles respectively in theories based on (affine extensions of) dual Lie algebras: for instance, between solitons of the $b_n^{(1)}$ theory and particles of the $c_n^{(1)}$ theory (and vice-versa), a case we shall discuss later. It also allows us to assign both particles and solitons unambiguously to spots on the diagram, and it may then be noticed that the soliton is obtained by folding a parent multisoliton whose soliton number is the order of the corresponding simple root under the automorphism. Explicit calculations of all the $\tau$-functions may be found in [27] whilst a general treatment in the formalism of Olive *et al.* may be found in [14].

Viewed in this way it is simple to compute a set of solutions $\delta\phi_{ab}$ to the Schrödinger problem (3.2). However, for the parent particle or soliton species which are not invariant under the folding, the construction of Olive *et al.* in [12] gives solutions in terms of $\hat{F}^k$ operators which are sums of such operators in the parent theory. This means that in any function $\delta\phi_{ab}$ for the non-simply-laced theories where $a$ or $b$ is one of these labels, the solution will contain two or more (possibly different) $X$ factors. However, the folding procedure ensures that the coefficients of the leading terms in $\delta\tau_j^{ab}$, which we denote by $X_{a\boldsymbol{b}}$, are in fact the same, and are easy to find for the non-simply-laced untwisted theories, since they come from a particle traversing the parent multisoliton. (To see this in terms of the full soliton solutions, see the $c_2^{(1)}$ calculation [16].) Another result is that, whereas for the simply-laced theories the $X$ factors (expressed as functions of $\cosh\eta$) were symmetric under the interchange of the soliton and particle labels, this is no longer the case. Since we now need to distinguish the soliton from the particle traversing it, we have put the soliton label in bold face; thus $X_{a\boldsymbol{b}}$ is the factor for particle $a$ traversing soliton $b$.

$c_n^{(1)}$, folded from $a_{2n-1}^{(1)}$ : a parent two-soliton ($a$ with $2n-a$) must be folded for $a = 1,..,n-1$, single for $n$; and so

$$\begin{aligned}
X_{b\boldsymbol{a}} &= X^P_{b\,a}X^P_{b\,2n-a} \\
X_{a\boldsymbol{n}} &= X^P_{a\,n} \\
X_{n\boldsymbol{a}} &= X^P_{n\,a}X^P_{n\,2n-a} \\
X_{n\boldsymbol{n}} &= X^P_{n\,n} \quad ,
\end{aligned}$$

where we have¡ used the superscript $P$ to denote an $X$ of the parent theory.

$b_n^{(1)}$, folded from $d_{n+1}^{(1)}$ : single for $1,..,n-1$, double for $n$;

$$\begin{aligned}
X_{b\boldsymbol{a}} &= X^P_{b\,a} \\
X_{n\boldsymbol{a}} &= X^P_{n\,a} \\
X_{a\boldsymbol{n}} &= X^P_{a\,n}X^P_{a\,n+1} \\
X_{n\boldsymbol{n}} &= X^P_{n\,n}X^P_{n\,n+1} \quad .
\end{aligned}$$

$g_2^{(1)}$, folded from $d_4^{(1)}$ : soliton 1 is a triple $(1+3+4)$ parent soliton, soliton 2 a single.

$$\begin{aligned}
X_{1\boldsymbol{1}} &= X^P_{1\,1}X^P_{1\,3}X^P_{1\,4} \\
X_{2\boldsymbol{1}} &= X^P_{2\,1}X^P_{2\,3}X^P_{2\,4} \\
X_{1\boldsymbol{2}} &= X^P_{1\,2} \\
X_{2\boldsymbol{2}} &= X^P_{2\,2} \quad .
\end{aligned}$$

Using these $X$s we can now go through the calculations described in the previous section. The only significant alteration to the method is in the condition for a discrete mode to have good asymptotic behaviour: (4.28) must be generalized.

- For a parent single $a$-soliton perturbing a parent multisoliton (of number $n$), the right-hand side must be replaced by $nm_b$.

  The $X_{a\cdot}$ factor is the same for each component of the multisoliton, and the bound state occurs when $X_{a\cdot} = 0$, when the leading term, with coefficient $(X_{a\cdot})^n$, vanishes. However, the next terms, which may be viewed as the highest terms in an $a$-soliton perturbing a parent $(n-r)$–soliton, have coefficients proportional to $(X_{a\cdot})^{n-r}$ and so also vanish. The first non-vanishing term behaves like $W_a/W_b^n$.

- Whenever the perturbing soliton is a parent multisoliton (4.28) remains valid. When the perturbed soliton is a parent single soliton this is straightforward.

  When both the perturbed and perturbing solitons are parent multisolitons, the highest term's coefficient is a product of $X$s coming from the different parent single-solitons. If one of these $X$s vanishes, then there will not be a zero in the others, and so the next-to-leading term in $\delta\phi^{ab}$ will not vanish. Thus the first non-vanishing term behaves like $W_a/W_b$.

It is possible that there could be further unexpected zeros in the (new) leading terms, which would affect our results. However, all such cases already have good asymptotic behaviour and so cannot be affected.

Using these asymptotics we obtain the results given in table 5, where in each case $\theta = \frac{\pi}{h}$.

Table 5: Non-simply-laced untwisted particle and soliton mass corrections

| | Particles | Solitons |
|---|---|---|
| $b_n^{(1)}$ | $\frac{m_a^{\rm qu}}{m_a^{\rm cl}} = \left(1 - \frac{\beta^2}{8h}\cot\theta + \frac{\beta^2}{4h^2} a\cot a\theta\right)$ | $\frac{M_a^{\rm qu}}{M_a^{\rm cl}} = \left(1 + \frac{\beta^2}{4\pi}\frac{h^\vee}{h} - \frac{\beta^2}{8h}\cot\theta + \frac{\beta^2}{4h^2} a\cot a\theta\right)$ <br> $\frac{M_n^{\rm qu}}{M_n^{\rm cl}} = \left(1 + \frac{\beta^2}{4\pi}\frac{h^\vee}{h}\right)$ |
| $c_n^{(1)}$ | $\frac{m_a^{\rm qu}}{m_a^{\rm cl}} = \left(1 - \frac{\beta^2}{16h}\cot\theta - \frac{\beta^2}{4h^2} a\cot a\theta\right)$ | $\frac{M_a^{\rm qu}}{M_a^{\rm cl}} = \left(\begin{array}{c} 1 + \frac{\beta^2}{4\pi}\frac{h^\vee}{h} - \frac{\beta^2}{16h}\cot\theta \\ + \frac{\beta^2}{4h^2}(n-a)\cot a\theta \end{array}\right)$ |
| $g_2^{(1)}$ | $\frac{m_1^{\rm qu}}{m_1^{\rm cl}} = \left(1 - \frac{7\beta^2}{144\sqrt{3}}\right)$ <br> $\frac{m_2^{\rm qu}}{m_2^{\rm cl}} = \left(1 - \frac{5\beta^2}{144\sqrt{3}}\right)$ | $\frac{M_1^{\rm qu}}{M_1^{\rm cl}} = \left(1 + \frac{\beta^2}{4\pi}\frac{h^\vee}{h} - \frac{\beta^2}{144\sqrt{3}}\right)$ <br> $\frac{M_2^{\rm qu}}{M_2^{\rm cl}} = \left(1 + \frac{\beta^2}{4\pi}\frac{h^\vee}{h} - \frac{5\beta^2}{144\sqrt{3}}\right)$ |

### 5.2.3 The $f_4^{(1)}$ results

The $f_4^{(1)}$ theory is obtained by folding the $e_6^{(1)}$ theory and the particle mass degeneracies are removed. We label the particles in accordance with the labelling of the particles of $e_6^{(1)}$ in [8] as $\{l, L, h, H\}$. They have masses

$$\begin{array}{rclcrcl} m_l^2 &=& (3-\sqrt{3})m^2 & \qquad & m_L^2 &=& 2(3-\sqrt{3})m^2 \\ m_h^2 &=& (3+\sqrt{3})m^2 & \qquad & m_H^2 &=& 2(3+\sqrt{3})m^2 \end{array}$$

We list the particle mass corrections below [24] for reference.

The solitons of types $l$ and $h$ are double solitons of the $e_6^{(1)}$ theories of types $\{l, \bar{l}\}$ and $\{h, \bar{h}\}$ respectively, while those of type $L$ and $H$ are single solitons of the parent theory. This gives the soliton masses as

$$\frac{M_l}{m_l} = \frac{M_h}{m_h} = -\frac{4h}{\beta^2}\,, \qquad \frac{M_L}{m_L} = \frac{M_H}{m_H} = -\frac{2h}{\beta^2}$$

We find the mass corrections (using Mathematica) as follows:

| | $m^{\rm qu}/m^{\rm cl}$ | $M^{\rm qu}/M^{\rm cl}$ |
|---|---|---|
| $l$ | $1-\frac{\beta^2}{384}(7+3\sqrt{3})$ | $1+\frac{3\beta^2}{16\pi}-\frac{\beta^2}{128}(1+\sqrt{3})$ |
| $L$ | $1-\frac{\beta^2}{384}(5+3\sqrt{3})$ | $1+\frac{3\beta^2}{16\pi}-\frac{\beta^2}{384}(5+3\sqrt{3})$ |
| $h$ | $1-\frac{\beta^2}{384}(3+5\sqrt{3})$ | $1+\frac{3\beta^2}{16\pi}-\frac{\beta^2}{384}(3+\sqrt{3})$ |
| $H$ | $1-\frac{\beta^2}{384}(9+\sqrt{3})$ | $1+\frac{3\beta^2}{16\pi}-\frac{\beta^2}{384}(9+\sqrt{3})$ |

### 5.2.4 Summary of non-simply-laced untwisted results

As before for the simply-laced and twisted theories, we find for the solitons which are single parent solitons the universal result (now with $k=1$)

$$\frac{M^{\rm qu}}{m^{\rm qu}} = -\frac{2h}{\beta^2}\left(1 + \frac{\beta^2 h^\vee}{4h\pi} \;+\; O(\beta^4)\right)\,.$$

# 6 Discussion

## 6.1 Comparison with exact S-matrices

These results have important implications for attempts to construct exact factorized $S$-matrices for the solitons. The preferred candidates for such $S$-matrices are the quantized affine algebra ($\mathrm{U}_q(\hat{g})$ ) invariant (trigonometric) solutions of the Yang-Baxter equation (YBE) corresponding to the *dual* $\hat{g}^\vee$ of the affine algebra $\hat{g}$ on which the Toda theory is based. The simply-laced affine algebras are self-dual, and Hollowood [18] has investigated the $a_n^{(1)}$ case. Apart from problems of unitarity, a fundamental problem with such $S$-matrices is that for $n > 2$ the topological charges of the (classical) solitons do not fill the fundamental representations of the Lie algebra [27]; and neither do those of the excited ('breathing') solitons [28]. It then has to be assumed that the quantum solitons can somehow take all the weights of each fundamental representation of $\mathrm{U}_q(\hat{g})$ (which is a reducible representation of $\mathrm{U}_q(g)$ whose highest component is the fundamental representation of $\mathrm{U}_q(g)$ ).

The non-simply-laced cases are even more subtle. For the twisted theories we would need to use YBE solutions based on untwisted algebras, which in the conventional approach have a rigid pole structure which respects the unrenormalized soliton mass ratios. For non-simply-laced untwisted theories the position is rather obscure, since very little is known about $R$-matrices for twisted algebras: as far as we are aware, only those for vector representations are known [29] and it is thus not clear whether the $R$-matrices in higher representations will follow the classical particle or soliton masses. The problem of rigid pole structure is likely to be resolved by making use of the 'spin gradation'[5], in which the fact that the non-local charges corresponding to different step operators have varying spins depending on the length of the associated root is apparently fundamental in ensuring crossing symmetry. Until recently this had only been investigated for the $a_2^{(2)}$ case [31], which is rather unfortunate since, as we have seen, $a_{2n}^{(2)}$ are the only non-simply-laced algebras for which the mass ratios remain constant. However, Babichenko [32] has now calculated the vector $a_5^{(2)}$ $S$-matrix, whose pole (with $m = 0$ in (37) ) reproduces precisely our $M_2^{qu}/M_1^{qu}$ ratio for $b_3^{(1)}$. The extension of this result to other cases is an important goal.

[5] We should like to thank D. Bernard for drawing our attention to this. See also [30]

## 6.2 Relating soliton and particle masses

It has been known for many years that it is consistent in sine-Gordon theory to identify the quantum particle with the first scalar breather state, and it is becoming apparent that the same is true of the other affine Toda theories, where for each soliton we expect the first scalar soliton-antisoliton bound state to be identified with the corresponding particle. The reason for this is unclear, but it is possible that one is simply calculating the first excited state in the zero topological charge sector by quantizing the theory in two equivalent ways: either by taking the free massive theory and quantizing the classical oscillator solutions, which must then be renormalized using the remaining, interactive, part of the Lagrangian; or by quantizing the classical breather solutions of the full theory. The latter, done for the sine-Gordon theory in [2], has not been carried out for the general case, and it remains an interesting problem to do so. However, we can make a tentative calculation of the breather mass by examining the breather pole in the soliton-antisoliton S-matrix. This was done for the vector soliton of $a_n^{(1)}$ by Hollowood [4], and for the other simply-laced algebras the expected S-matrix pole gives us

$$m^{\rm qu}_{\rm particle} = m^{\rm qu}_{\rm first\ breather} \;=\; -2M^{\rm qu}_{\rm soliton}\sin\left(\frac{\pi B}{2h}\right)\;. \tag{6.1}$$

This also works for the $a_2^{(2)}$ model considered by Smirnov [33] (when we make the connection $\beta^2 = 4\gamma$ with his notation).

So, for the simply-laced solitons, we have found the perturbative formulae in table 3, which are consistent with a conjectured exact formula (6.1).

### 6.2.1 Twisted theories

For each species of soliton in the twisted theories we also find a close relationship between the particle and soliton masses, as given by eqn. (5.1),

$$M^{\rm qu}_a = -\frac{h^\vee}{\pi B}\, m^{\rm qu}_a \;\;+\;\; O(\beta^2)\;. \tag{6.2}$$

Thus the ratios of soliton masses are equal to the ratios of particle masses in these theories. If we were to identify the first breather with the particle, then we could conjecture an exact formula

$$m^{\rm qu}_a = -2M^{\rm qu}_a \sin\left(\frac{\pi B}{2h^\vee}\right)\;. \tag{6.3}$$

### 6.2.2 Non-simply-laced untwisted theories

In the remaining cases, that is non-simply-laced untwisted theories, we find that for those soliton species which derive from a single soliton in the parent theory, there is the same relationship between the particle and soliton masses (6.2) as in the simply-laced and twisted theories, and we could again conjecture an exact formula (6.3).

For the solitons which derive from a $n$-fold parent multiple soliton, there are *extra* quantum corrections which reduce the mass of this soliton below the expected result,

$$M_a^{\rm qu} = -\, n \, \frac{h^\vee}{\pi B} \, m_a^{\rm qu} \, \Big( 1 + \beta^2 \epsilon \Big) \;\; + \;\; O(\beta^2) \; ,$$

where $\epsilon$ is positive and $\beta^2$ is negative.

In the cases of both affine and Lie duality, since we expect to be able to identify the particle with the first breather, a better understanding of how the masses are related will come with attempts to construct exact S-matrices and understand their pole structure.

## 6.3 Duality

### 6.3.1 Affine duality

The subject of duality in real-coupling theories has been much discussed [6, 9, 17, 36], and both mass ratios and S-matrices have been related in theories based on dual affine algebras (we recall that dual affine Lie algebras are related by changing the directions of the arrows on their affine Dynkin diagrams). For dual theories $g^{(k)}$ and $g'^{(k')}$ the coupling constants are related by $\beta^2 \beta'^2 = 16\pi^2 k k'$ : the weak-coupling regime of one can be identified with the strong-coupling regime of the other, and we can interpolate between the theories.

In sine-Gordon theory, and in all the simply-laced theories as well where $k = k' = 1$, we expect that the theory is in fact only well defined for $\beta^2 > -4\pi$, and so this duality will relate a well defined theory to an undefined theory (as it will for the non-simply-laced theories as well.) Thus we feel that we cannot at present offer any useful ideas on affine duality in the imaginary coupling theories.

### 6.3.2 Lie duality

The results of Olive *et al.* [1, 12–15] suggested the presence of some sort of 'Lie duality' which relates the solitons in one untwisted theory $(g)^{(1)}$ to the particles in the theory $(g^\vee)^{(1)}$ based on the dual of the finite Lie algebra. This is exemplified by the pair $b_n^{(1)}$ and $c_n^{(1)}$, for which Olive *et al.* found the classical results

$$\frac{M_a(b_n^{(1)})}{m_a(c_n^{(1)})} = \frac{M_a(c_n^{(1)})}{m_a(b_n^{(1)})} = -\frac{2\sqrt{2}h}{\beta^2}\ , \tag{6.4}$$

where the $\beta$ dependence only occurs in the soliton masses – the particle masses are $\beta$ independent. One might hope that this would extend in some way to the quantum theory, i.e. that we might be able to find some functions $\lambda(\beta)$, $\tilde{\lambda}(\beta)$ and $\beta'(\beta)$ such that

$$M_a(b_n^{(1)})\Big|_\beta \ = \ \lambda(\beta)\, m_a(c_n^{(1)})\Big|_{\beta'}\ , \tag{6.5}$$

$$m_a(b_n^{(1)})\Big|_\beta \ = \ \tilde{\lambda}(\beta)\, M_a(c_n^{(1)})\Big|_{\beta'}\ . \tag{6.6}$$

The strongest form of duality one could think of would be one in which $\lambda = \tilde{\lambda}$. From the 'classical'[6] result (6.4) this would relate the coupling constants by

$$\beta^2\beta'^2 = 8h$$

and so would, like the 'affine duality' above, relate strong and weak coupling regimes of different theories. Immediately, this implies that if we wish to use small coupling $\beta$ results we must replace the $\beta'$ dependent quantities in (6.5) and (6.6) by the quantum results at strong coupling, which we clearly cannot do. Since we can only calculate weak coupling expansions, we cannot test such a duality directly, but we can point out an example of quantum behaviour which indicates that any such duality cannot be implemented in a simple fashion.

Let us suppose that there is a sense in which strong-weak duality is valid. We can look at, for example, the classical $\beta \to 0$ limit of the ratio

$$\lim_{\beta\to 0} \frac{M_a^{\rm qu}\,/\,m_a^{\rm qu}}{M_n^{\rm qu}\,/\,m_n^{\rm qu}}(b_n^{(1)}) = \frac{M_a^{\rm cl}\,/\,m_a^{\rm cl}}{M_n^{\rm cl}\,/\,m_n^{\rm cl}}(b_n^{(1)}) = \frac{1}{2}\ . \tag{6.7}$$

The corresponding quantity at strong $\beta$ could be calculated in the $c_n^{(1)}$ theory, and we find

$$\frac{m_a^{\rm cl}\,/\,M_a^{\rm cl}}{m_n^{\rm cl}\,/\,M_n^{\rm cl}}(c_n^{(1)}) = \frac{1}{2}\ . \tag{6.8}$$

[6] n.b. there are no particles in the classical theory: the existence of particles is a purely quantum phenomenon and the particle masses are purely quantum in origin, containing a factor of $\hbar$, set here to 1.

If we believe that there is some duality relating the two quantities (6.7) and (6.8), then since it is the same for zero $\beta$ and zero $\beta'$, it is hard to reconcile this with the fact that these two quantities have very different behaviours when we include the quantum corrections in $\beta$ and $\beta'$:

$$\frac{M_a^{\rm qu}/m_a^{\rm qu}}{M_n^{\rm qu}/m_n^{\rm qu}}(b_n^{(1)}) = \frac{1}{2}\left(1-\frac{\beta^2}{8h}\cot\theta\right) , \tag{6.9}$$

$$\frac{m_a^{\rm qu}/M_a^{\rm qu}}{m_n^{\rm qu}/M_n^{\rm qu}}(c_n^{(1)}) = \frac{1}{2}\left(1-n\frac{\beta'^2}{4h^2}\cot(a\theta)\right) . \tag{6.10}$$

This argument, while not at all ruling out the possibility of a 'strong-weak' duality, does mean that it cannot be implemented in any particularly simple fashion.

If we instead postulate a 'weak-weak' duality, between the weak-coupling regime in one theory and the weak-coupling regime in the other, then we can try to keep (6.5) and (6.6) but with different $\lambda, \tilde{\lambda}$. This was in fact the remarkable coincidence observed in $c_2^{(1)}$ in [16]. However, if we demand $\beta = \beta'$, then this is only true for $c_2^{(1)}$, and even if we allow $\beta^2 = A\beta'^2(1+O(\beta'^2))$ for some constants $A$ we still cannot satisfy (6.5) or (6.6) for general $n$.

Overall we must therefore say that, despite no conclusive proof of the absence of such a result, we find no evidence of Lie duality.

### 6.4 Unitarity

As pointed out in [16], our calculations remain in some sense formal, since the Hamiltonian is complex. However, for the classical soliton solutions not only the energy [12] but also an infinite number of conserved quantities [34] is real, so there is hope that we might be able to restrict the fields or the Hilbert space to the solitons and their bound states, so that all observables take real values and a quantum theory can be defined. In this spirit it seems significant to us that all our calculations have been done within the multisoliton sector, *i.e.* that the complete set of solutions to the Schrödinger problem (3.2) is found by examining multisolitons.

### 6.5 Boundary state independence

Recall that in section 3 we imposed periodic boundary conditions on the perturbation of the soliton to be quantized. Although sine-Gordon theory has been considered before, it is only in the last year or so that careful consideration has been given to the question of which boundary conditions may be imposed on Toda theories without destroying integrability (see ref. [35] and refs. therein), and we believe that the condition we impose does not in fact preserve this integrability. However, we expect that, since the period is eventually taken to infinity, our results will not be affected in any case.

**Acknowledgments**

NJM would like to thank E. Corrigan, A. Koubek, A. J. Macfarlane, N. Manton, T. Samols and P. Sutcliffe for discussions and P. E. Dorey and H. Braden for helpful communications. GMTW would like to thank D. Bernard, A. Koubek, D. I. Olive and J.-B. Zuber for helpful conversations, comments and criticism at various stages. NJM was supported by a PPARC fellowship. GMTW was supported by a EU HCM grant.

## References

[1] D. I. Olive, N. Turok and J. W. R. Underwood,
*Affine Toda solitons and vertex operators*, Nucl. Phys. B409 (1993) 509–546.

[2] R. F. Dashen, B. Hasslacher and A. Neveu, Phys. Rev. D10 (1974) 4114; Phys. Rev. D11 (1975) 3424; Phys. Rev. D12 (1975) 2443.

[3] R. Rajaraman, *Solitons and Instantons: an introduction to solitons and instantons in quantum field theory*, North-Holland, 1982.

[4] T. J. Hollowood,
*Quantum soliton mass corrections in $SL(N)$ affine Toda field theory*, Phys. Lett. B300 (1993) 73–83, hep-th/9209024.

[5] B. L. Feigin and E. V. Frenkel,
*Integrals of motion and quantum groups,* YITP/K-1036, hep-th/9310022, To appear in the proceedings of the CIME Summer school on integrable systems and quantum groups, Montecatini Terme, Italy, 14-22 Jun 1993, hep-th/9310022;
G. W. Delius, M. T. Grisaru and D. Zanon,
*Quantum conserved currents in affine Toda theories*, Nucl. Phys. B385 (1992) 307–328, hep-th/9202069

[6] H. G. Kausch and G. M. T. Watts,
*Duality in quantum Toda theory and W algebras*, Nucl. Phys. B386 (1992) 166–192, hep-th/9202070.

[7] A. B. Zamolodchikov and Al. B. Zamolodchikov,
*Factorized S-matrices in two dimensions as the exact solution of certain relativistic quantum field theory models*, Ann. Phys. (NY) 120 (1979) 253.

[8] H. Braden, E. Corrigan, P. E. Dorey and R. Sasaki,
*Extended Toda field theory and exact S-matrices,* Phys. Lett. B227 (1989) 411;
*Affine Toda field theory and exact S matrices,* Nucl. Phys. B338 (1990) 689–746;
*Multiple poles and other features of affine Toda field theory*, Nucl. Phys. B356 (1991) 469–497.

[9] G. W. Delius, M. T. Grisaru and D. Zanon,
*Exact S-matrices for the non-simply-laced affine Toda theories $a^{(2)}_{2n-1}$*, Phys. Lett. B277 (1992) 414, hep-th/9112007;
*Exact S-Matrices for non-simply-laced affine Toda theories*, Nucl. Phys. B382 (1992) 365–408, hep-th/9201067.

[10] P. Christe and G. Mussardo,
*Integrable systems away from criticality: the Toda field theory and S matrix of the tricritical Ising model*, Nucl. Phys. B330 (1990) 465;
*Elastic S matrices in (1+1)-dimensions and Toda field theories*, Int. J. Mod. Phys. A5 (1990) 4581–4628.

[11] T. J. Hollowood,
*Solitons in affine Toda field theories*, Nucl. Phys. B384 (1992) 523–540.

[12] D. I. Olive, N. Turok and J. W. R. Underwood,
*Solitons and the energy momentum tensor for affine Toda theory*, Nucl. Phys. B401 (1993) 663–697.

[13] M. A. C. Kneipp and D. I. Olive,
*Crossing and anti-solitons in affine Toda theories*, Nucl. Phys. B408 (1993) 565–578, hep-th/9305160.

[14] M. A. C. Kneipp and D. I. Olive,
*Solitons and vertex operators in twisted affine Toda field theories*, University College, Swansea, Preprint SWAT-93-94-19, hep-th/9404030.

[15] A. Fring, P. R. Johnson, M. A. C. Kneipp and D. I. Olive,
*Vertex operators and soliton time delays in affine Toda field theory*, University College, Swansea, Preprint SWAT-93-94-30, hep-th/9405034.

[16] G. M. T. Watts,
*Quantum mass corrections for* $C_2^{(1)}$ *affine Toda theory solitons*, Phys. Lett. B338 (1994) 40, hep-th/9404065.

[17] E. Corrigan, P.E. Dorey and R. Sasaki,
*On a generalized bootstrap principle*, Nucl. Phys. B408 (1993) 579–599;
H. S. Cho, I. G. Koh and J. D. Kim,
*Duality in the* $D_4$ *affine Toda theory,* Phys. Rev. D47 (1993) 2625–2628;
G. M. T. Watts and R. A. Weston,
$G_2^{(1)}$ *affine Toda field theory: a numerical test of exact S matrix results*, Phys. Lett. B289 (1992) 61–66.

[18] T. J. Hollowood,
*Quantizing SL(N) solitons and the Hecke algebra*, Int. J. Mod. Phys. A8 (1993) 947–982.

[19] P. E. Dorey,
*Root systems and purely elastic S-matrices*, Nucl. Phys. B358 (1991) 654–676.

[20] P. E. Dorey,
*Root systems and purely elastic S matrices. 2*, Nucl. Phys. B374 (1992) 741–762.

[21] P. E. Dorey,
*Partition functions, intertwiners and the Coxeter element*, Int. J. Mod. Phys. A8 (1993) 193–208.

[22] A. Fring and D. I. Olive,
*The fusing rule and the scattering matrix of affine Toda theory*, Nucl. Phys. B379 (1992) 429–447.

[23] H. Braden,
*A note on affine Toda couplings*, J. Math. Phys A25 (1992) L15–L20.

[24] H. Braden, *private communication.*

[25] D. I. Olive and N. Turok,
*Local conserved densities and zero-curvature conditions for Toda lattice field theories*, Nucl. Phys. B257 [FS14] (1985) 277–301.

[26] N. J. MacKay and W. A. McGhee,
*Affine Toda solitons and automorphisms of Dynkin diagrams*, Int. J. Mod. Phys. A8 (1993) 2791–2807, Erratum Int. J. Mod. Phys. A8 (1993) 3830; hep-th/9208057.

[27] W. A. McGhee, *On the topological charges of the affine Toda theory solitons*, PhD thesis, Durham University, 1994;
*The topological charges of the* $a_n^{(1)}$ *affine Toda solitons*, Int. J. Mod. Phys. A9 (1994) 2645–2666, hep-th/9307035.

[28] U. Harder, A. Iskandar and W. A. McGhee,
*On the breathers of* $a_n^{(1)}$ *affine Toda field theory*, Durham preprint DTP-94-35, hep-th/9409035.

[29] M. Jimbo,
*Quantum R-matrix for the generalized Toda system*, Comm. Math. Phys. 102 (1986) 537.

[30] A. J. Bracken, G. W. Delius, M. D. Gould and Y.-Z. Zhang,
*Infinite families of gauge-equivalent R-matrices and gradations of quantized Affine algebras*, Int. J. Mod. Phys. A (1994), hep-th/9310183.

[31] C. J. Efthimiou,
*Quantum group symmetry for the $\phi_{12}$-perturbed and $\phi_{21}$-perturbed minimal models of conformal field theory*, Nucl. Phys. B398 (1993) 697–740.

[32] A. Babichenko,
*Integrable vector perturbations of W-invariant theories and their quantum group symmetry*, Racah Inst. preprint RI-7, hep-th/9406197, revised 18/11/94..

[33] F. A. Smirnov,
*Exact S matrices for $\phi_{12}$-perturbed minimal models of conformal field theory*, Int. J. Mod. Phys. A6 (1991) 1407–1428.

[34] M. D. Freeman,
*Conserved charges and soliton solutions in affine Toda theory*, King's College London preprint KCL-TH-94-8, hep-th/9408092.

[35] E. Corrigan, P. E. Dorey, R. H. Rietdijk and R. Sasaki,
*Affine Toda field theory on a half-line*, Phys. Lett. B333 (1994) 83–91, hep-th/9404108 ;
E. Corrigan, P. E. Dorey and R. H. Rietdijk,
*Aspects of affine Toda field theory on a half-line*, DTP-94-29, hep-th/9407148.

[36] P. E. Dorey,
*A Remark on the coupling dependence in affine Toda field theories*, Phys. Lett. B312 (1993) 291–298, hep-th/9304149.

[37] A. Fring, H. C. Liao and D. I. Olive,
*The Mass spectrum and coupling in affine Toda theories*, Phys. Lett. B266 (1991) 82–86.

# A Appendix: Comparison of the conventions of Dorey with those of Olive et al.

In this appendix we summarize the results of Fring and Olive ([FO]), [22] and Dorey ([D]) [20, 21] which we need. We confine ourselves to simply-laced cases.

Both [FO] and [D] use simple roots $\alpha_a$ and split these into two groups, black and white. They define fundamental Weyl reflections $w_a$ and composite reflections

$$w_{\{\bullet\}} = \sigma_- = \prod_{\text{black}} w_a \ \ , \ w_{\{\circ\}} = \sigma_+ = \prod_{\text{white}} w_a \ .$$

and give a distinguished Coxeter element

$$w = \sigma = w_{\{\bullet\}} w_{\{\circ\}} = \sigma_- \sigma_+ \ .$$

Both [FO] and [D] give a set of representatives of the orbits of the Coxeter element. These are

| | | $\bullet$ | $\circ$ |
|---|---|---|---|
| [FO] | $\gamma_i$ | $-\alpha_i$ | $\alpha_i$ |
| [D] | $\phi_a$ | $w_{\{\circ\}}\alpha_a$ | $\alpha_a$ |

with the result that these are related by $\phi_a = -w_{\{\circ\}}\gamma_j$. Since

$$w w_{\{\bullet\}} = w_{\{\bullet\}} w^{-1} \ ,$$

they find the relation

$$\gamma_a \cdot \sigma^p \gamma_b = \phi_a \cdot w^{-p} \phi_b = (\phi_a, w^{-p}\phi_b) \ .$$

They also define integers related to the black and white sets by

| | | $\bullet$ | $\circ$ |
|---|---|---|---|
| [FO] | $c(i)$ | $-1$ | $1$ |
| [D] | $u_a$ | $0$ | $1$ |

with

$$u_{ab} \equiv u_a - u_b = \frac{c_a - c_b}{2} \ .$$

which allows them to relate the orbit representatives for conjugate particles by

$$\phi_a = -w^{\frac{h+u_{a\bar{a}}}{2}} \phi_{\bar{a}} \ \ , \ \gamma_{\bar{j}} = -\sigma^{-\frac{h}{2}+\frac{c_j - c_{\bar{j}}}{4}} \gamma_j \ . \qquad \text{(A.1)}$$

[FO] define eigenvectors $q(s)$ of the Coxeter element, with eigenvalues $\exp 2\pi \mathrm{i} s/h$, where $s$ is an exponent. These are normalized

$$|q(s)|^2 = h \ \ , \ \ q(s)^* = q(h-s) \ .$$

[D] defines eigenvectors of the Cartan matrix, $q_j^{(s)}$, where $s$ is an exponent and $j$ goes from 1 to $r$ and labels the root. These are normalized

$$|q^{(s)}|^2 = 1 \ \ , \ \ q_\bullet^{(s)} = q_\bullet^{(h-s)} \ \ , \ \ q_\circ^{(s)} = -q_\circ^{(h-s)} \ .$$

We can relate these two by:

$$q(s)\cdot\gamma_j = -ie^{-i\theta_s(1-u_j)}\sqrt{2h}\, q_j^{(s)} \ ,$$

with the result that

$$|q(1)\cdot\gamma_j| = \sqrt{2h}\, q_j^{(1)} \ .$$

This gives the particle masses for the simply-laced theories as

$$m_a = m|\gamma_a \cdot q(1)| = \left(m\sqrt{2h}\right) q_a^{(1)} \ . \tag{A.2}$$

An important identity[7] comes from the Fourier transform of eqn. (2.15) of ref. [21].

$$\textstyle\sum_{p=0}^{h-1} \cos(\frac{\pi}{h}(2p + u_{ab}))(\phi_a, w^{-p}\phi_b) = 2h\, q_a^{(1)} q_b^{(1)} \ . \tag{A.3}$$

We also use the identities

$$\left(m\sqrt{2h}\right) \phi_\circ \cdot w^{-p} a_\circ^1 = 2m_\circ \cos 2p\theta \ \ , \ \left(m\sqrt{2h}\right) \phi_\circ \cdot w^{-p} w_{\{\circ\}} a_\bullet^1 = 2m_\circ \cos(2p+1)\theta \ . \tag{A.4}$$

To derive these, remember (in the notation of [D]) that two eigenvectors of the Coxeter element are

$$(l_\circ^1 - \mathrm{e}^{-\mathrm{i}\theta} l_\bullet^1) \ , \ \ (l_\circ^1 - \mathrm{e}^{\mathrm{i}\theta} l_\bullet^1) \ ,$$

with eigenvalues $\exp \mathrm{i}\theta$ and $\exp -\mathrm{i}\theta$ respectively, where $\theta = \pi/h$. Hence, with

$$a_\circ^1 = (l_\circ^1 - \mathrm{e}^{-\mathrm{i}\theta} l_\bullet^1) + (l_\circ^1 - \mathrm{e}^{\mathrm{i}\theta} l_\bullet^1) \ , \ \ a_\bullet^1 = (l_\bullet^1 - \mathrm{e}^{\mathrm{i}\theta} l_\circ^1) + (l_\bullet^1 - \mathrm{e}^{-\mathrm{i}\theta} l_\circ^1) \ ,$$

and that

$$\sum_\circ m_\circ \phi_\circ = m\sqrt{2h} a_\circ^1 \ \ , \ \sum_\bullet m_\bullet \phi_\bullet = m\sqrt{2h} w_{\{\circ\}} a_\bullet^1 \ .$$

[7]We should like to thank P. Dorey for this result

## B Appendix: Simply-laced real coupling mass corrections

We wish to work out the 1-loop mass corrections to a simply-laced theory. Let us restrict to the case $h$ even and a root labelling such that $u_a = 1$, i.e. $\phi_a$ is 'white'. We have

$$\delta m_a^2 = \sum_{(b,c)\to a} -\frac{C_{abc}^2}{4\pi\delta} \tan^{-1}_{[0,\pi)} \left( \frac{|\delta|}{m_b^2 + m_c^2 - m_a^2} \right) ,$$

where the sum is over all ordered pairs of particles $(b, c)$ which have three point couplings to $a$. From [8] we have

$$|C_{abc}| = \frac{4\beta\Delta_{abc}}{\sqrt{h}} , \quad \Delta_{abc} = \frac{1}{2} m_b m_c \sin \bar{u}_{bc}^a .$$

where, (as ever), $0 \le \bar{u}_{bc}^a < \pi$. Using $m_a^2 = m_b^2 + m_c^2 - 2m_b m_c \cos \bar{u}_{bc}^a$, we find

$$\delta m_a^2 = -\frac{\beta^2}{2h\pi} f_a , \quad f_a = \sum_{(b,c)\to a} m_b m_c \bar{u}_{bc}^a \sin \bar{u}_{bc}^a .$$

Let us consider simply $f_a$, which we can rewrite using the symmetric property of fusion as

$$f_a = \sum_{(a,b)\to c} 2m_a m_b (\frac{\pi}{2} - \bar{u}_{ab}^c) \sin \bar{u}_{ab}^c . \tag{B.1}$$

From Dorey, [19], there is a simple relation between the values which $\phi_a \cdot w^{-p} \phi_b$ takes and the possibility of a fusion $(a, b) \to ?$. We give these relations in table 6.

Table 6: Table of fusions

| $\phi_a \cdot w^{-p}\phi_b$ | 0 | 1 | 2 | −1 | −2 |
|---|---|---|---|---|---|
| fusion | — | $a \times \bar{b}$ | — | $a \times b$ | — |

Let us consider the cases ±1 in turn:

### B.1 $\phi_a \cdot w^{-p}\phi_b = -1$

In this case, there exists a fusion $a \times b \to \bar{c}$, for some $b$ and $c$, and a relation amongst roots of the form

$$\phi_a + w^{-p}\phi_b + w^{-q}\phi_c = 0 .$$

From this we can deduce the value of $\bar{u}_{ab}^c$,

| $p$ | $\bar{u}_{ab}^c$ |
|---|---|
| $0 \dots \frac{h}{2} - 1$ | $\pi - (2p + u_{ab})\theta$ |
| $h/2 \dots h - 1$ | $(2p + u_{ab})\theta - \pi$ |

where as usual $\theta = \pi/h$ and since $u_a = 1$, $u_{ab}$ can take the values $0, 1$ only.

Then the sum over allowed fusions can be restricted to a sum over the values of $p$ and $b$ for which $\phi_a \cdot w^{-p}\phi_b = -1$, to give, remembering that each fusion occurs exactly twice in this list ( see [37] ),

$$f_a = -m_a \sum_b \sum_{p:\phi_a\cdot w^{-p}\phi_b=-1} m_b \left[ \sum_{p=0}^{h-1} [\frac{\pi}{2} - (2p+u_{ab})\theta] \sin(2p+u_{ab})\theta + \sum_{p=h/2}^{h-1} \pi \sin(2p+u_{ab})\theta \right] .$$

## B.2 $\phi_a \cdot w^{-p}\phi_b = +1$

These cases also give us twice the exact amount of fusions, but in this case of the form $a \times \bar{b} \to \bar{c}$. Now, if we use eqn. (A.1) to relate the representatives for conjugate particles, we find

| $p$ | $\bar{u}^c_{a\bar{b}}$ |
|---|---|
| $0 \ldots \frac{h}{2} - 1$ | $(2p+u_{ab})\theta$ |
| $h/2 \ldots h-1$ | $2\pi - (2p+u_{ab})\theta$ |

and so we can again write $f_a$ in terms of these cases, but this time as

$$f_a = m_a \sum_b \sum_{p:\phi_a\cdot w^{-p}\phi_b=+1} m_b \left[ \sum_{p=0}^{h-1} [\frac{\pi}{2} - (2p+u_{ab})\theta] \sin(2p+u_{ab})\theta + \sum_{p=h/2}^{h-1} \pi \sin(2p+u_{ab})\theta \right] .$$

If we observe that the terms for $\phi_a \cdot w^{-p}\phi_b = \pm 2$ give zero, we can now add together the two expressions for $f_a$ to get

$$f_a = \sum_b \frac{1}{2} m_a m_b \left[ \begin{array}{c} \sum_{p=0}^{h-1} (\phi_a\cdot w^{-p}\phi_b)[\frac{\pi}{2} - (2p+u_{ab})\theta] \sin(2p+u_{ab})\theta \\ + \sum_{p=h/2}^{h-1} (\phi_a\cdot w^{-p}\phi_b) \pi \sin(2p+u_{ab})\theta \end{array} \right] .$$

We can now sum over black and white separately, use the results (A.4) and recombine the black and white terms to get

$$\begin{aligned} f_a &= m_a^2 \left[ \sum_{r=1}^{2h-1} \left(\frac{\pi}{2} - r\theta\right) \sin r\theta \cos r\theta + \sum_{r=h}^{2h-1} \pi \sin r\theta \cos r\theta \right] \\ &= \frac{1}{2} m_a^2 \pi \cot \frac{\pi}{h} \, , \end{aligned} \qquad \text{(B.2)}$$

as required, and the simply-laced one-loop mass formula is proven for white roots and $h$ even.

We leave the case of $h$ odd to the reader!

# C Appendix: Exceptional series $X$ factors

Throughout this appendix, we use the conventions that

$$(\pm x) = \cosh\eta \mp \cos\frac{x\pi}{h}\ , \quad \{\pm x\} = \frac{\cosh\eta \mp \cos\frac{x\pi}{h}}{\cosh\eta \pm \cos\frac{x\pi}{h}}$$

Table 7: $e_6^{(1)}$ X factors

| | $l$ | $\bar{l}$ | $L$ | $\bar{h}$ | $h$ | $H$ |
|---|---|---|---|---|---|---|
| $l$ | $\frac{(+0)(+6)}{(-4)(+2)}$ | $\frac{(-2)(+4)}{(-0)(+6)}$ | $\frac{(+3)(-5)}{(-3)(+5)}$ | $\frac{(+1)(+5)}{(-3)(+3)}$ | $\frac{(+3)(-3)}{(-1)(-5)}$ | $\frac{(+2)}{(-2)}$ |
| $\bar{l}$ | | $\frac{(+0)(+6)}{(-4)(+2)}$ | $\frac{(+3)(-5)}{(-3)(+5)}$ | $\frac{(+3)(-3)}{(-1)(-5)}$ | $\frac{(+1)(+5)}{(+3)(-3)}$ | $\frac{(+2)}{(-2)}$ |
| $L$ | | | $\frac{(+0)(-2)(+4)}{(-0)(+2)(-4)}$ | $\frac{(+2)}{(-2)}$ | $\frac{(+2)}{(-2)}$ | $\frac{(+1)(+5)}{(-1)(-5)}$ |
| $\bar{h}$ | | | | $\frac{(+0)(+6)}{(-2)(-4)}$ | $\frac{(+2)(+4)}{(-0)(+6)}$ | $\frac{(+1)(+3)}{(-1)(-3)}$ |
| $h$ | | | | | $\frac{(+0)(+6)}{(-2)(-4)}$ | $\frac{(+1)(+3)}{(-1)(-3)}$ |
| $H$ | | | | | | $\frac{(+0)(+2)(+4)}{(-0)(-2)(-4)}$ |

Table 8: $e_7^{(1)}$ X factors

| | 1 | 2 | 3 | 4 |
|---|---|---|---|---|
| 1 | {+0}{−2}{+8} | {+5}{−7} | {+4}{−6}{+8} | {+1}{−3}{+7} |
| 2 | | {+0}{−2}{+6}{−8} | {+3}{−5}{+7} | {+4}{−8} |
| 3 | | | {+0}{−2}{+4} | {+3} |
| 4 | | | | {+0}{−4}{+6}{+8} |
| | | 5 | 6 | 7 |
| 1 | | {+4}{−8} | {+2}{−4}{+6} | {+3} |
| 2 | | {+1}{−3}{+5} | {+3} | {+2}{+8} |
| 3 | | {+2}{+8} | {+2}{+6}{+8} | {+1}{+5} |
| 4 | | {+3}{+5}{−7} | {+1}{+7} | {+2}{+4} |
| 5 | | {+0}{+6} | {+2}{+4} | {+1}{+3}{+7} |
| 6 | | | {+0}{+4}{+8} | {+1}{+3}{+5} |
| 7 | | | | {+0}{+2}{+4}{+6} |

Table 9: $e_8^{(1)}$ X factors

| | 1 | 2 | 3 | 4 |
|---|---|---|---|---|
| 1 | {+0}{−2}{+10}{−12} | {+6}{−8}{+12}{−14} | {+1}{−3}{+9}{−13} | {+5}{−7}{+9}{−11}{+13} |
| 2 | | {+0}{−2}{+6}{−8}{−14} | {+5}{−9}{+11} | {+3}{−5}{+7} |
| 3 | | | {+0}{−4}{+8}{−12}{−14} | {+4}{+14} |
| 4 | | | | {+0}{−2}{+4}{+10}{−12}{+14} |
| | 5 | 6 | 7 | 8 |
| 1 | {+2}{+4}{+8}{−14} | {+5}{−9}{+11} | {+3}{−5}{+7} | {+4}{+14} |
| 2 | {+4}{+14} | {+1}{−3}{+5}{+11}{−13} | {+3}{+9}{−11}{+13} | {+2}{+8} |
| 3 | {+1}{−5}{+7}{+9}{−13} | {+4}{+6}{−8}{+12} | {+2}{+8} | {+3}{+5}{−13} |
| 4 | {+3}{+7}{−9}{+11} | {+2}{+8} | {+2}{+6}{+12}{−14} | {+1}{+5}{+9} |
| 5 | {+0}{+9}{+10}{−12} | {+3}{+5}{+13} | {+1}{+5}{+9} | {+2}{+4}{+6}{+12} |
| 6 | | {+0}{+6}{+10}{−14} | {+2}{+4}{+8}{+14} | {+1}{+3}{+7}{+9} |
| 7 | | | {+0}{+4}{+6}{+10} | {+1}{+3}{+5}{+7}{+11} |
| 8 | | | | {+0}{+2}{+4}{+6}{+8}{+10} |